\UseRawInputEncoding
\documentclass[12pt,letterpaper]{article}
\usepackage{amsmath,amssymb,pgf,pgfarrows,pgfnodes,float,appendix, hyperref}
\usepackage{graphicx}
\usepackage{subfigure}
\usepackage[margin=0.9in]{geometry}

\newcommand{\be}{\begin{equation}}
\newcommand{\ee}{\end{equation}}
\newcommand{\bea}{\begin{eqnarray}}
\newcommand{\eea}{\end{eqnarray}}

\title{{\rm\footnotesize \qquad \qquad \qquad \qquad \qquad \ \qquad \qquad \qquad \ \ \ \ \ \            UTTG-13-2022      RUNHETC-2022-35}\vskip.5in   Discretely Charged Dark Matter in Inflation Models Based on Holographic Space-time}
\author{Tom Banks\\
Department of Physics and NHETC\\
Rutgers University, Piscataway, NJ 08854\\
E-mail: \href{mailto:tibanks@ucsc.edu}{tibanks@ucsc.edu}
\\
\\
Willy Fischler\\
Department of Physics and Texas Cosmology Center\\
University of Texas, Austin, TX 78712\\
E-mail: \href{mailto:fischler@physics.utexas.edu}{fischler@physics.utexas.edu}
\\
\\
}
\date{}
\begin{document}
\maketitle

\begin{abstract} The Holographic Space-time (HST) model of inflation has a potential explanation for dark matter as tiny primordial black holes.   Motivated by a recent paper of Barrau\cite{barrau} we propose a version of this model where some of the Inflationary Black Holes (IBHs), whose decay gives rise to the Hot Big Bang, carry the smallest value of a discrete symmetry charge.  The fraction $f$ of IBHs carrying this charge is difficult to estimate from first principles, but we fix it by requiring that the crossover between radiation and matter domination occurs at the correct temperature
$T_{eq} \sim 1 eV = 10^{-28} M_P$.  The fraction is small, $f \sim 2\times 10^{-9}$ so we believe this gives an extremely plausible model of dark matter.  \end{abstract}
\maketitle

\section{Introduction}

\begin{figure}[H]
\centering
\includegraphics[scale=0.3]{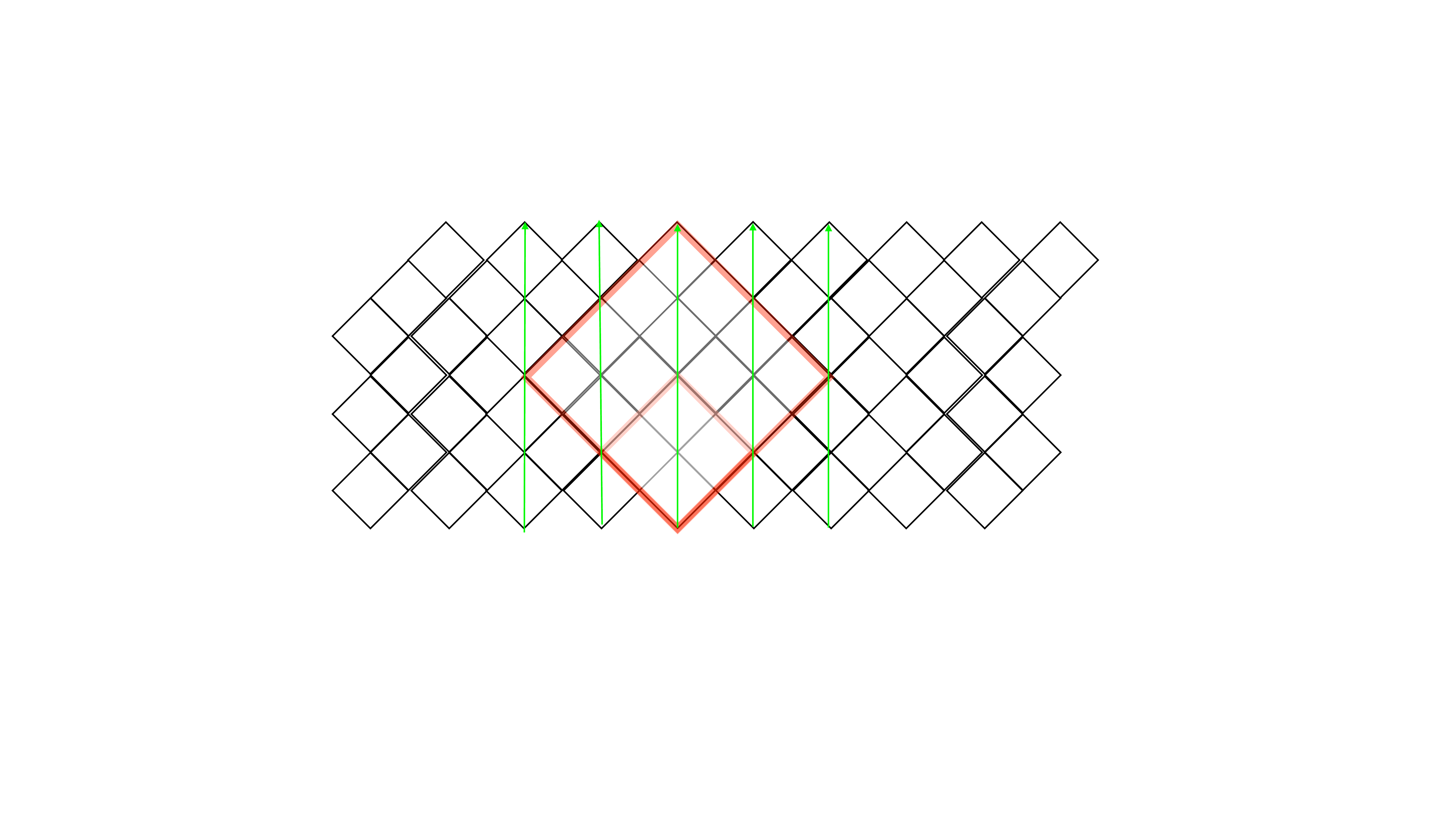}
\caption{Holographic Inflationary Cosmology in Conformal Time: Equal Time Surfaces are Hyperbolae Interpolating Between Diamonds}
\label{holoinflation}
\end{figure}

The HST model of inflation\cite{holoinflation} is a finite quantum mechanical model, which gives a very economical explanation of known facts about the very early universe.   HST models are based on Jacobson's Principle: the Einstein equations are the hydrodynamic equations of the area law $S = A/4G$ applied to any causal diamond in a Lorentzian space-time.  Therefore one should search for quantum models whose hydrodynamics agrees with some particular solution of Einstein's equations. The features of HST inflation models are easily summarized
\begin{itemize}

\item The model consists of a large number of independent quantum systems, describing the universe as viewed from different geodesics in an FRW space-time.
The relation between proper time and area of the holographic screen of a diamond with past tip on the singular beginning of the universe, is matched to the relationship between the time in the quantum theory and the entropy of the density matrix assigned to the diamond.
\item The Hamiltonian is time dependent, in order to ensure that degrees of freedom inside a given causal diamond form an independent subsystem.  This also provides a natural resolution of the Big Bang singularity: when the Hilbert space of a diamond is small enough the hydrodynamic description breaks down but the quantum mechanics is well defined and finite.
\item A particular soluble model, in which for each proper time $t$, the modular Hamiltonian of a diamond is the $L_0$ generator of a cutoff conformal field theory on 
an interval of length $I$ with a UV cutoff $l$, such that $I/l \gg 1$ (but $t$ independent) and central charge scaling like $t^2$, is "dual" to a flat FRW geometry with scale factor \begin{equation} a(t) = \sinh^{1/3} (3t/R_I) . \end{equation}  These models have no localized excitations and saturate the covariant entropy bound at all times.
\item Inflationary models are obtained by insisting that the dynamics follow the soluble model for a large number of e-folds ($80$ is what seems to fit the data of our universe) after which the diamond Hilbert space slowly expands so that it can fit $> e^{80}$ copies of the original space.  What one would have called gauge copies of the causal diamond in a de Sitter space with radius $R_I$ become localized excitations of the expanded diamond, with all of the statistical properties of black holes of radius $\sim R$.  See Figure\ref{holoinflation} for a cartoon of how this happens. This gives rise to a novel theory of CMB fluctuations, with $\zeta \sim (R_I \epsilon)^{-1}$ and the scalar to tensor ratio $r \sim \epsilon^{-2}$.   Properties of spinning black holes and the $1 +1$ CFT model of the horizon pin down the coefficients in these relations, in which $\epsilon = -\frac{\dot{H}}{H^2}$.  One requires a different slow roll metric than conventional inflationary models, to fit the data on the CMB.

\item Evaporation of the "inflationary black holes" (IBHs) gives rise to the hot Big Bang and baryogenesis.  
\item The only element of very early universe cosmology that is not explained simply by the model is Dark Matter.  We have speculated\cite{previouspbh} that mergers of the tiny IBHs might form a collection of Primordial Black Holes (PBHs) consistent with astronomical data\cite{Carretal}.  
\end{itemize} 

Recently, Barrau\cite{barrau} has argued that the merger scenario cannot work, but that a model in which evaporation of the IBHs left over Planck scale remnants, could explain dark matter.  While we remain a bit skeptical that one can come to Barrau's negative conclusion without dedicated computer simulations, we found the idea of remnants intriguing.  GellMann's Totalitarian Principle is an axiomatization of a known fact about quantum systems.  Transition matrix elements exist unless some (approximate) conservation law forbids them.  Put simply, Planck mass black hole remnants cannot be ruled out by Hawking's thermodynamic arguments, but they are implausible unless there is some quantum number that prevents them from decaying.  

In models of quantum gravity, charges carried by local excitations are always coupled to gauge fields.  The most innocuous kind of gauge field, from the point of view of a dark matter model is associated with a discrete gauge group like $Z_N$.   In gauge theories, charged particles always experience long range interactions.  For discrete gauge theories, these are just Aharonov Bohm interactions with topological cosmic strings so $Z_N$ charged particles will behave like neutral dark matter.

The theory of supersymmetry (SUSY) breaking contains a natural discrete gauge symmetry in a universe like our own with small positive cosmological constant.  If one imagines the model of our universe to be part of a (possibly discrete) family of models that converges to zero c.c., then the limiting model is likely to be supersymmetric.  This is a "phenomenological" observation.  There are no perturbative string models in Minkowski space that violate SUSY.  There are no known sequences of $AdS/CFT$ models with tunable c.c., which violate SUSY in the limit\footnote{SUSY violating relevant perturbations of SUSic models represent large objects embedded in AdS space.  Physics far from the center becomes exactly SUSic.} of small c.c..   

On the other hand generic SUGRA Lagrangians with SUSY preserving minima have negative c.c. because of a term proportional to the square of the superpotential in the vacuum energy.  One of us\cite{tbR} pointed out long ago that the criterion for a SUSic vacuum with vanishing c.c. was preservation of an R symmetry.   The R symmetry must be discrete\cite{komseib}.     

The R symmetry acts chirally on the gravitino, and keeps it massless, but in dS space there are processes where the R symmetry is broken by absorption and re-emission of gravitinos at the horizon.  In\cite{cosmosusybreak} it was postulated that R symmetry violating terms in the low energy effective Lagrangian, induced by this non-local effect, would trigger the super-Higgs effect in a self consistent manner.  This leads to an equation for the gravitino mass in terms of the c.c.,
\begin{equation} m_{3/2} = K \Lambda^{1/4} , \end{equation} where it has proven difficult to estimate the constant $K$.  

A discrete R symmetry and a light gravitino does not at first sound like a recipe for obtaining stable R charged black hole remnants.   The remnants can emit gravitinos and reduce their R charge.   However, there are many examples in field theory where the lowest charge under some discrete gauge group is carried by a very heavy particle.
This means that the symmetry breaking induced by the effective gravitino mass leaves over a discrete subgroup of the high energy discrete gauge symmetry.  The only instability of the heavy R charged black holes will be moving through the horizon, or spontaneous nucleation of a black hole of opposite charge, which is a highly improbable process.  

\section{Phenomenology of Discretely Charged PBH Dark Matter}  

In the context of the HST model of inflation, it is simple to incorporate discrete charges that stabilize a fraction $f$ of IBHs at the Planck scale.   Inflation is followed by an early matter dominated era, in which the matter is composed of IBHs.  For comparison we can calculate the expected fraction of magnetically charged black holes using the black hole entropy formula, according to which the expected fraction of black holes in a random sample is
\begin{equation} f = e^{-\frac{Q^2M_P^2}{ \alpha M^2}} , \end{equation} where $Q$ is the integer valued magnetic charge and $\alpha$ is the value of the fine structure constant at the scale of the Schwarzschild radius. Taking $M \sim M_P$, $Q = \pm 1$ and $\alpha$ equal to its value at the scale of unification of standard model couplings this gives $f \sim 10^{-12.5}$.  There are many issues with this estimate, the most serious of which is using a formula from statistical mechanics for a low entropy system, but it gives us a general sense that $f$ should be small but not doubly exponentially small.  

In HST inflation models, the number density of IBHs at the end of inflation is
\begin{equation} n_{IBH} \sim C R_I^{-3} , \end{equation} where $C \sim 1/30$ is the minimal dilution factor necessary to assure that the IBHs do not immediately coalesce to form a maximum entropy state.  The inflationary Hubble radius, $R_I$ in Planck units, is determined by matching to the size of CMB fluctuations
\begin{equation} R_I = \epsilon^{-1} 10^{5} , \end{equation} and $\epsilon$ is bounded by the requirement that slow roll expansion is faster than fast scrambling of the black holes.  This again follows from the requirement that black holes remain isolated quantum subsystems during the slow roll era.  The bound is 
\begin{equation} \epsilon > ({\rm ln}\ R_I)^{-1} . \end{equation}  This is satisfied for $\epsilon \sim 0.1$ .   We insist on being close to the bound because we require the highest probability of initial conditions that lead to a universe with localized excitations.  Since the power spectrum of CMB fluctuations in these models scales like $\epsilon^{-2}$, this value is roughly consistent with data.  

The universe remains matter dominated until a time $t_D = 10240 \pi  g^{-1} R_I^3$, when most of the IBHs decay into radiation.  $g$ is the number of particle species below the Hawking temperature of the IBHs.  The resulting energy density of radiation at the beginning of the Hot Big Bang is
\begin{equation} \rho_{\gamma} = \frac{C}{R_I^2} t_D^{-2} =  C g^2 (10240 \pi)^{-2} R_I^{-8} = \frac{\pi^2}{30} g T_{RH}^4. \end{equation}
If a fraction $f$ of the IBHs leave over Planck scale remnants, then their energy density at reheating is
\begin{equation} \rho_{rem} = f C R_I^{-3} t_D^{-2}, \end{equation} and
\begin{equation} \frac{\rho_{rem}}{\rho_{\gamma}} = f R_I^{-1} . \end{equation}
This ratio grows like $\frac{T_{RH}}{T}$ as the radiation gas cools, and hits $1$ when
\begin{equation} T_{eq} = (R_I)^{-1} f T_{RH}  = 10^{-28} , \end{equation} where the last equality is the observed temperature at which matter radiation crossover occurs. 
$T_{RH}$ is given by 
\begin{equation} \approx \frac{g^{1/4}}{100 \pi R_I^2} \sim  0.5\times 10^{-13} .\end{equation} 
Thus
\begin{equation} f \approx 2 \times 10^{-9}. \end{equation} 

So, what appears to be a reasonable estimate of the probability of discretely charged black holes being formed in HST models is consistent with the data.

One more issue needs to be addressed.  In previous publications\cite{previouspbh} we've argued that if a fraction $f \sim 10^{-24}$ of black holes of mass $\sim 10^{11}$ are formed during the early matter dominated era, then these could account for the observed value of the matter radiation crossover.   In the present scenario, these are unnecessary, and could even become an embarrassment.   Since these PBHs are not cosmologically stable, their decay could lead to signatures that have been ruled out by observation.  It is possible that during the matter dominated era below $T = 10^{-28}$ the unstable PBHs merge into more stable ones before too much Hawking radiation has been emitted.  Ongoing computer simulations will determine whether this is plausible\cite{anish}.   If it is, we will have two competing models that account for the data.  It seems highly unlikely that the probabilities work out so that both contributions to dark matter have comparable densities, but if they do one could have a scenario where some of the dark matter decays.  Such models have been invoked to explain some of the apparent discrepancies between data and the LCDM model.   From the present point of view, the simplest idea is that the fraction of merged IBHs which could survive down to $T = 10^{-28}$ is negligible, and that discretely charged dark matter (DCDM) accounts for everything we see.

\section{Conclusions} 

Motivated by a suggestion of Barrau, we propose that HST inflation models incorporate 
a discrete $Z_N$ gauge symmetry, and that a fraction $f \sim 10^{-9}$ of the erstwhile inflationary horizon volumes in the model, carry the smallest value of $Z_N$ charge.  This discrete symmetry group could be the remnant of a larger discrete R symmetry, broken by gravitino interactions with the horizon, which generate the gravitino mass.  The resulting models account, at the order of magnitude level, with everything we know about the cosmology of the very early universe.  Inflation ends in an early matter dominated era, dominated by IBHs with Schwarzschild radii approximately equal to the inflationary horizon size.  Most of the IBHs decay, producing the Hot Big Bang and baryogenesis.  Those charged under $Z_N$ become the dark matter.  And the rest is history...

\end{document}